\documentclass[manuscript,screen]{acmart}
\AtBeginDocument{%
  }

\setcopyright{acmlicensed}
\usepackage{enumitem}
\copyrightyear{2018}
\acmYear{2018}
\acmDOI{XXXXXXX.XXXXXXX}
\acmConference[Conference acronym 'XX]{Make sure to enter the correct
  conference title from your rights confirmation email}{June 03--05,
  2018}{Woodstock, NY}
\acmISBN{978-1-4503-XXXX-X/2018/06}

\begin{document}

\title{Security Analysis of ChatGPT: Threats and Privacy Risks}

\author{Yushan Xiang}
\affiliation{
  \institution{Hainan University}
  \country{China}
}
\authornote{Both Yushan Xiang and Zhongwen Li are co-first authors.}
\author{Zhongwen Li}
\authornotemark[1]
\affiliation{%
  \institution{Hainan University}
  \country{China}
}
 \email{lzw123@hainanu.edu.cn}

\author{Xiaoqi Li}
\affiliation{
  \institution{ Hainan University}
  \country{China}
}
\email{csxqli@ieee.org}


\begin{abstract}
  As artificial intelligence technology continues to advance, chatbots are becoming increasingly powerful. Among them, ChatGPT, launched by OpenAI, has garnered widespread attention globally due to its powerful natural language processing capabilities based on the GPT model, which enables it to engage in natural conversations with users, understand various forms of linguistic expressions, and generate useful information and suggestions. However, as its application scope expands, user demand grows, and malicious attacks related to it become increasingly frequent, the security threats and privacy risks faced by ChatGPT are gradually coming to the forefront.
In this paper, the security of ChatGPT is mainly studied from two aspects, security threats and privacy risks. The article systematically analyzes various types of vulnerabilities involved in the above two types of problems and their causes. Briefly, we discuss the controversies that ChatGPT may cause at the ethical and moral levels. In addition, this paper reproduces several network attack and defense test scenarios by simulating the attacker's perspective and methodology. Simultaneously, it explores the feasibility of using ChatGPT for security vulnerability detection and security tool generation from the defender's perspective.

\end{abstract}



\maketitle

\section{Introduction}
\

Artificial Intelligence (AI), as a key technology to equip computer systems with learning and reasoning capabilities, has experienced several rounds of development since it was proposed in the mid-20th century. From the early artificial neural networks, the Turing test, and expert systems, to the rise of deep learning and large language models in recent years, AI technology continues to promote the process of social intelligence\cite{gallifant2024peer}. In particular, the Transformer model proposed by Google in 2017 laid the foundation for the rapid evolution of Large Language Models (LLMs)\cite{min2023recent}. Since then, representative models such as BERT, GPT, etc. have demonstrated excellent performance in natural language processing, speech recognition, image generation, and other fields. Currently, major domestic technology enterprises have also launched their large models, such as DeepSeek, Doubao, kimi, etc, to promote the vigorous development of the AI industry.

ChatGPT, as a representative generative AI model launched by OpenAI, has been widely used in many fields such as text writing, code generation, knowledge quizzes, etc., with its powerful language generation capability and good user interaction experience, bringing significant convenience to society. However, at the same time, its risks in terms of cybersecurity, privacy protection, copyright compliance, and educational ethics have gradually emerged \cite {saharia2022photorealistic}. Studies have shown that ChatGPT may be abused for writing malicious code, generating false information, bypassing security restrictions, etc., while there are privacy issues such as training data leakage, model poisoning, and information reconstruction. Some studies at home and abroad have paid attention to ChatGPT-related attack methods and abusive behaviors, such as prompt injection, phishing email generation, model stealing, etc., and called for strengthening the prudent regulation of its use\cite{takale2024cyber}.

In this context, this paper centers on the security threats and privacy risks of ChatGPT. Firstly, we sort out the development background of big language models and the technical principles of ChatGPT, and analyze its typical applications in multiple practical scenarios. Subsequently, we summarize and categorize the main security issues and privacy risks faced by ChatGPT, such as malicious content generation, training data leakage, model reverse engineering, etc. Finally, through simulated attack and defense experiments, we validate the feasibility of the relevant attack modes, and explore their potential countermeasures and improvement suggestions\cite{charfeddine2024chatgpt}.
\section{Background}
\subsection{ChatGPT}
\

Since OpenAI first released ChatGPT in 2018, its model architecture and capabilities have gone through several rounds of upgrading and evolution, driving the widespread application of big language models in natural language processing\cite{lucchi2024chatgpt}. It was not until the end of 2022 that OpenAI launched the ChatGPT-3.5 version, which truly introduced the dialogue mechanism and was widely promoted in an open form, becoming a key node in the drive to massify generative AI\cite{pan2020privacy}.
\begin{itemize}
    \item  \textbf{ChatGPT-1(2018):} ChatGPT-1 is the first version based on the Transformer architecture, adopting a structure containing only a Decoder Block with 12 layers and 117M parameters. The model performs text categorization and generation tasks, and has preliminary Q \& A capabilities\cite{ruixue2023insights}. Through a combination of generative pre-training and discriminative fine-tuning, ChatGPT-1 shows excellent performance in some natural language processing tasks. Nevertheless, due to the lack of conversational capability and limited generalization, it does not yet have the characteristics of a true conversational system.

\item  \textbf{ChatGPT-2 (2019):} ChatGPT-2 is significantly extended in architecture with 48 layers and 1542M parameters. The model migrated to a variety of supervised tasks through unsupervised training and achieved leading performance in 7 out of 8 language modeling benchmarks\cite{radford2019language}. Compared to its predecessor, ChatGPT-2 demonstrates enhanced language understanding and text generation capabilities, enabling simple conversations, story creation, and has even been used to generate content such as phishing emails, showing potential risk of abuse\cite{el2023integration}.

\item  \textbf{ChatGPT-3 (2020):} A leap in scale and capability, with the number of parameters raised to 175B, a 96-layer structure, and a large model based entirely on self-supervised learning. It can accomplish multiple tasks without fine-tuning, which significantly improves the quality of language generation and multi-task adaptation\cite{radford2018improving}. However, its training process cannot effectively identify harmful information in network data, which may lead to biased or ethical issues in the generated content. In addition, ChatGPT-3 still does not have a true multi-round natural dialog function\cite{adhikari2024exploring}.

\item \textbf{ChatGPT-4 (March 2023):} ChatGPT-4 is the latest upgraded version at present, and although the parameter scales and architectural details have not been fully disclosed, several evaluations have shown that it significantly outperforms version 3.5 in academic and professional exams\cite{brown2020language}. Although ChatGPT-4 is mainly offered as a paid service, its improvements in language comprehension, reasoning ability, and response stability make it more suitable for demanding application scenarios\cite{abdaljaleel2024multinational}.

\end{itemize}

ChatGPT, as an advanced conversational AI model, has been widely used in customer service, writing, translation, and programming. In customer service scenarios, it can efficiently answer user questions and reduce enterprise labor costs\cite{briganti2024chatgpt}. In the writing and media industry, it can assist in content generation and information refinement, improving creative efficiency. In translation, ChatGPT supports personalized adjustment of tone and style. In programming practice, it can be used for code debugging, annotation, and generation, significantly improving development efficiency\cite{achiam2023gpt}.
\subsection{Transformer}
\

\begin{figure}[H]
    \centering
    \includegraphics[width=0.5\linewidth]{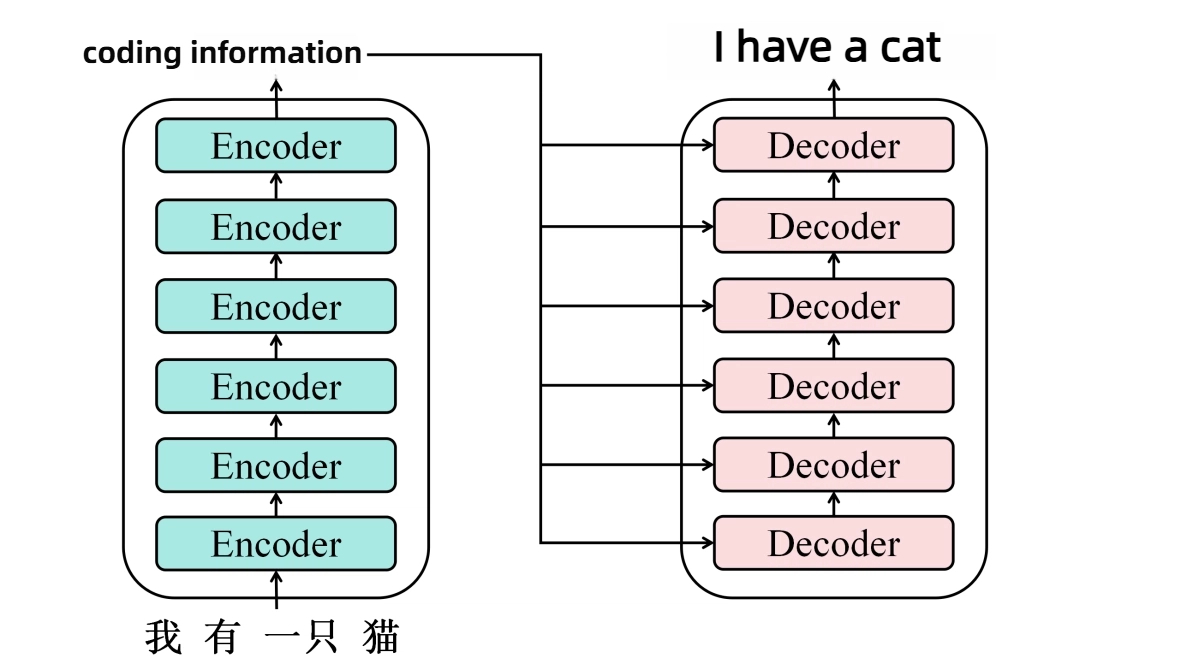}
    \caption{Overall Structure of the Transformer Model}
    \label{fig:1}
\end{figure}

As illustrated in  Fig. \ref{fig:1}, the Transformer architecture is composed of an Encoder and a Decoder, each comprising six blocks.
\begin{itemize}
    \item  Step 1: 
As shown in Fig. \ref{fig:2}, the model sums the word vectors (word embedding) of each word with its corresponding position vectors (position embedding) in the input stage to obtain the input representation matrix of each word. The word vectors represent the mapping of words into a high-dimensional vector space to capture their semantic features\cite{ashish2017attention}. In contrast, the position vectors are used to encode information about the order of words in a sentence, compensating for the model's inability to perceive word order.
Where n denotes the number of words in the sentence and d denotes the dimension of the vector. With this embedding approach, the model is able to incorporate positional information while retaining semantic information, thus understanding and processing natural language text more effectively\cite{jiang2024transformer}.
\begin{figure}[H]
    \centering
    \includegraphics[width=0.5\linewidth]{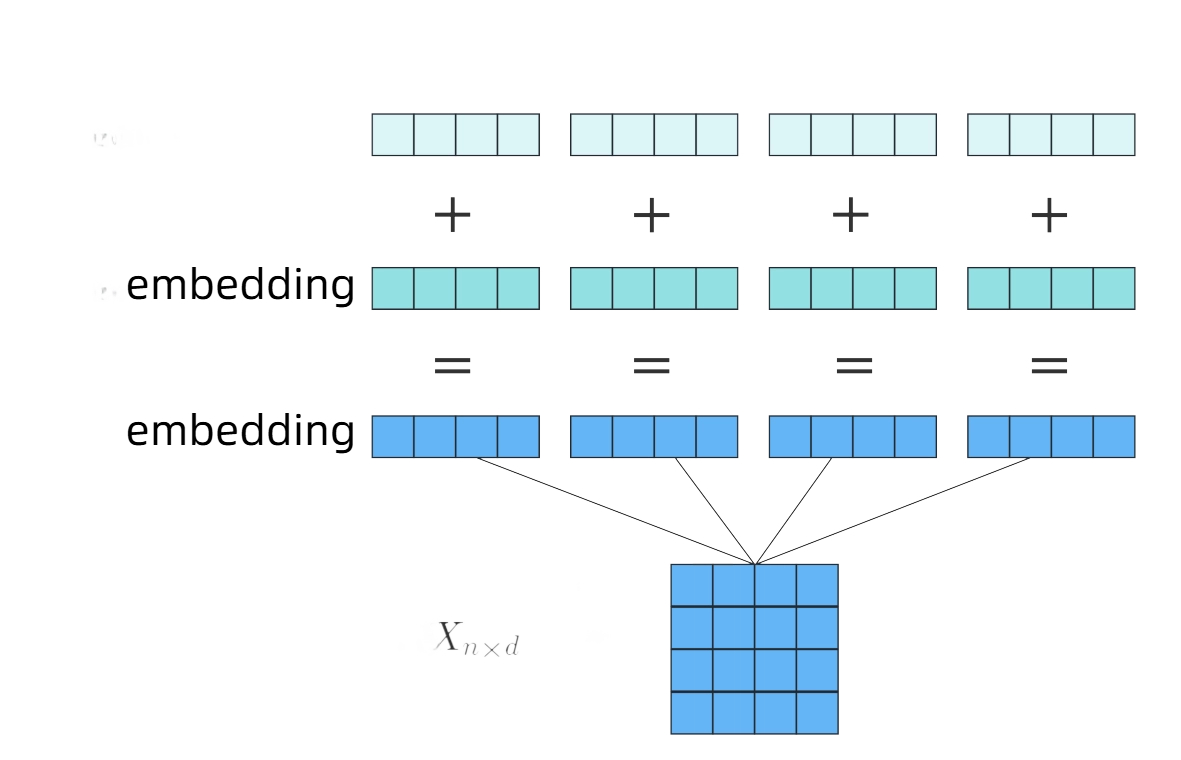}
    \caption{Transfomer Model - Embedding Layer}
    \label{fig:2}
\end{figure}
\item  Step 2: As shown in Fig. \ref{fig:3}, the vector matrix of words is input into the Encoder, and after successive layer-by-layer processing of 6 layers of Encoder Blocks, an encoded matrix containing semantic information of all the words in the sentence is finally output\cite{zhao2024dynamic}.
\begin{figure}[H]
    \centering
    \includegraphics[width=0.5\linewidth]{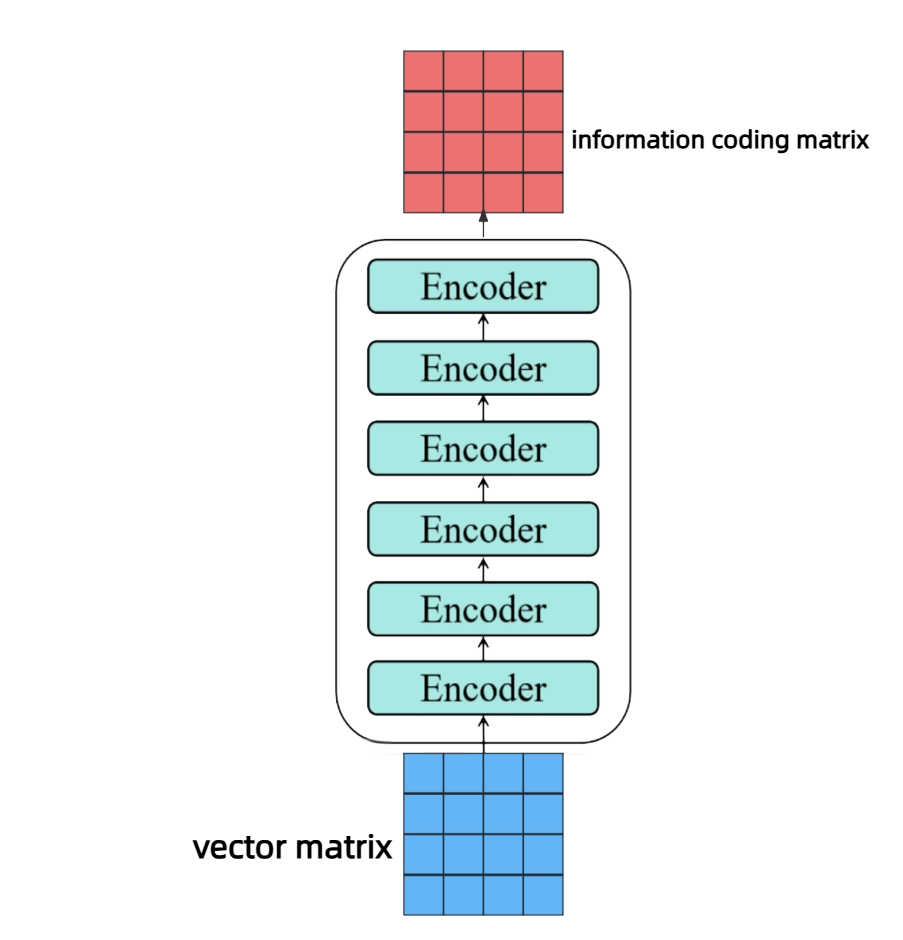}
    \caption{Transfomer Model-Encoder Information Coding}
    \label{fig:3}
\end{figure}
\item  Step 3: As shown in Fig. \ref{fig:4}, when the Decoder generates the $(i+1)$ word, it can only attend to the first $i$ previously generated tokens. To enforce this constraint, the Transformer model employs a Masked Multi-Head Attention mechanism, which prevents the model from accessing future positions during decoding\cite{fui2023generative}. The term "masked" indicates that the attention vector at each position is computed solely based on the current and previous tokens, thereby avoiding information leakage from subsequent words\cite{tang2024survey}.
\begin{figure}[H]
    \centering
    \includegraphics[width=0.5\linewidth]{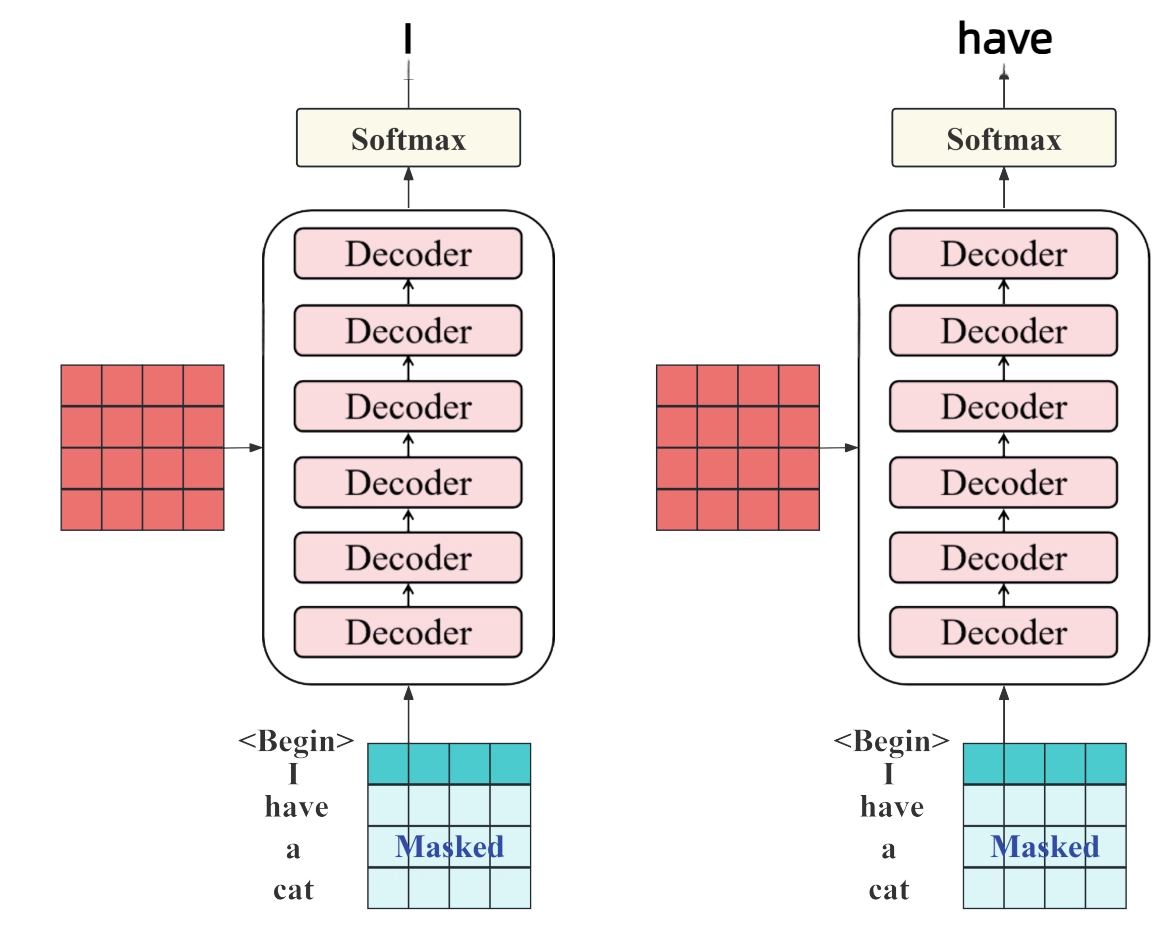}
    \caption{ Transfomer Model-Decoder Prediction}
    \label{fig:4}
\end{figure}

\end{itemize}

\subsection{Input}
\

Word embeddings are typically pre-trained on large-scale corpora such as Wikipedia, GloVe, Word2Vec, and FastText. These models map each word into a high-dimensional vector space, capturing semantic relationships among words, and providing rich representations for downstream tasks.
Unlike RNNs or CNNs, the Transformer model lacks an inherent mechanism for modeling word order\cite{sallam2023chatgpt}. Therefore, it incorporates Positional Encoding to inject explicit position information into the input representations. Positional encoding vectors have the same dimensionality as word embeddings and vary based on the word's position in the sentence\cite{li2024transformer}.
In the original Transformer implementation proposed by Vaswani et al., positional encodings are computed using sinusoidal functions as follows in equation \ref{eq:1}and equation \ref{eq:2}.$pos$ is the position index of the word in the sentence (starting from 0),$i$ is the dimension index of the positional encoding, and $d$ is the total dimensionality (equal to the word embedding dimension). This method offers two main advantages.After computing the word embedding $x_i$ and the positional encoding $p_i$, the final input representation for each token is obtained by element-wise addition. The full input matrix for a sentence can be expressed as equation \ref{eq:4}.  $E$ is the input embedding matrix,
 $X$ is the word embedding matrix,
$P$ is the positional encoding matrix,
 $n$ is the number of tokens in the sentence,
   $d$ is the embedding dimension\cite{hou2024conv2former}.

\begin{equation}
\text{PE}_{(pos, 2i)} = \sin\left(\frac{pos}{10000^{2i/d}}\right)
\label{eq:1}
\end{equation}
\begin{equation}
\text{PE}_{(pos, 2i+1)} = \cos\left(\frac{pos}{10000^{2i/d}}\right)
\label{eq:2}
\end{equation}

\begin{enumerate}
    \item \textbf{Scalability}: The encoding is not limited by sentence length seen during training. It can generalize to longer sequences by computing encodings for unseen positions.
    \item \textbf{Relative position modeling}: Due to the periodic nature of sine and cosine functions, the model can more easily capture relative distances between tokens\cite{hatamizadeh2025mambavision}.
    \[
    \sin(A + B) = \sin A \cos B + \cos A \sin B
    \]
    \[
    \cos(A + B) = \cos A \cos B - \sin A \sin B
    \]
\end{enumerate}

\begin{equation}
e_i = x_i + p_i
\label{eq:3}
\end{equation}

\begin{equation}
E = X + P \in \mathbb{R}^{n \times d}
\label{eq:4}
\end{equation}

\subsection{Self-Attention Mechanism}
\

As shown in Figure~\ref{fig:5}, the Transformer model consists of two main components. The Encoder on the left and the Decoder on the right. Within the architecture, there are three {Multi-Head Attention (MHA) modules in total, each composed of multiple Self-Attention mechanisms\cite{zheng2024enhanced}. In the Encoder, there is a single multi-head attention layer. In contrast, the Decoder includes two Multi-Head Attention layers, one of which employs Masked Attention to prevent positions from attending to subsequent positions during training, ensuring the autoregressive property of language generation\cite{huang2020metapoison}.
After each Multi-Head Attention operation, two essential components are applied.
\begin{figure}[H]
    \centering
    \includegraphics[width=0.3\linewidth]{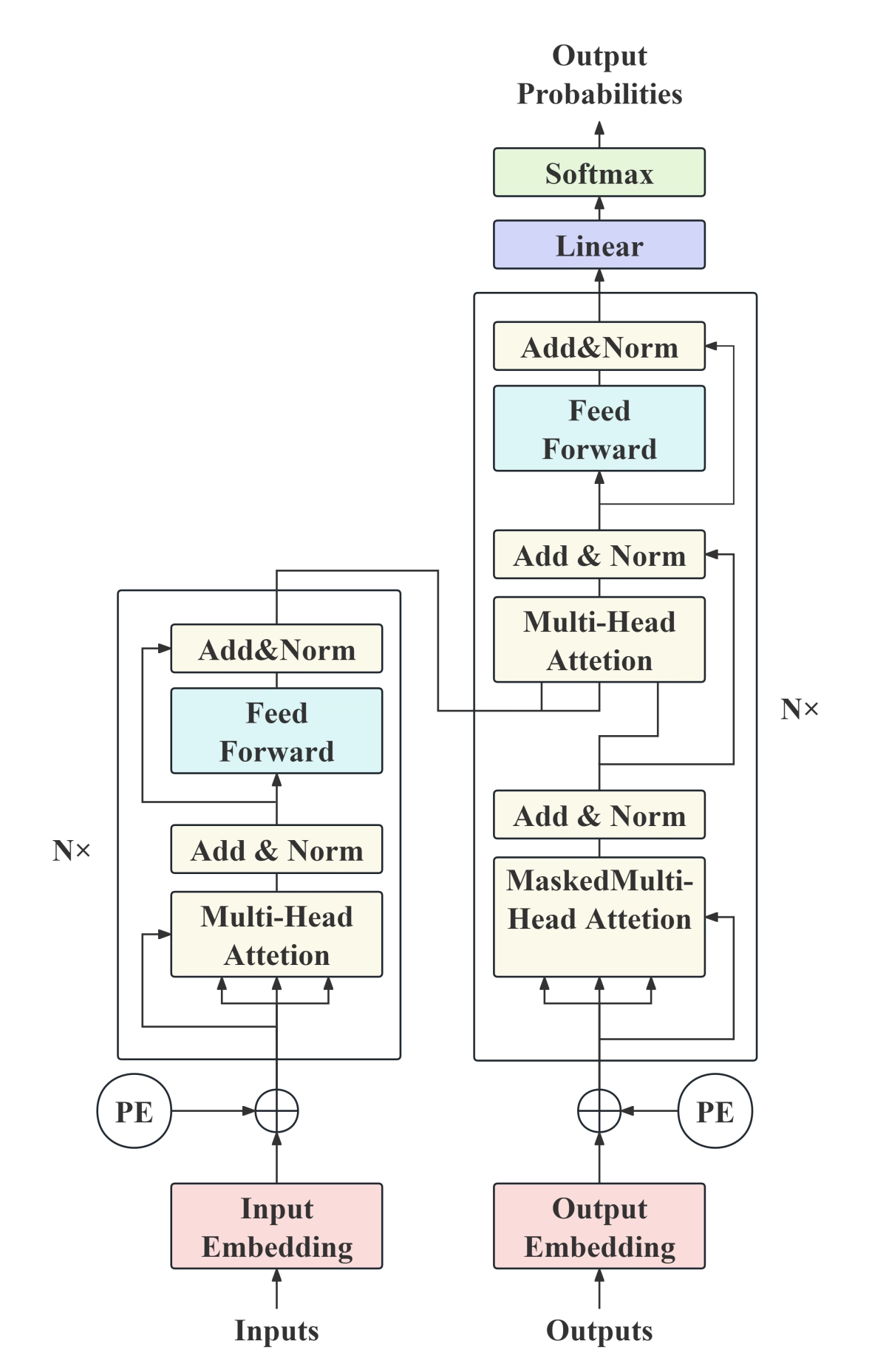}
    \caption{Encoder and Decoder Specific Architecture}
    \label{fig:5}
\end{figure}

\begin{itemize}
    \item \textbf{Add}: This represents a Residual Connection, which helps mitigate the vanishing gradient problem and improves gradient flow through the network.
    \item \textbf{Norm}: This refers to Layer Normalization, which standardizes the activations within each layer, stabilizing and accelerating training.
\end{itemize}

As shown in Figure~\ref{fig:6}, the input to the self-attention mechanism is a vector matrix $\mathbf{X}$, which can either be the initial word embeddings or the output of a previous encoder block\cite{shumailov2021sponge}. To compute attention scores, the input matrix $\mathbf{X}$ is linearly projected into three different representations.the query matrix $\mathbf{Q}$, the key matrix $\mathbf{K}$, and the value matrix $\mathbf{V}$. These are calculated as follows in equation \ref{eq:5}.$\mathbf{W}^Q$, $\mathbf{W}^K$, and $\mathbf{W}^V$ are learnable weight matrices used to transform the input into the respective spaces\cite{meng2024refined}.

\begin{equation}
\mathbf{Q} = \mathbf{X} \mathbf{W}^Q, \quad
\mathbf{K} = \mathbf{X} \mathbf{W}^K, \quad
\mathbf{V} = \mathbf{X} \mathbf{W}^V
\label{eq:5}
\end{equation}

\begin{figure}[H]
    \centering
    \includegraphics[width=0.5\linewidth]{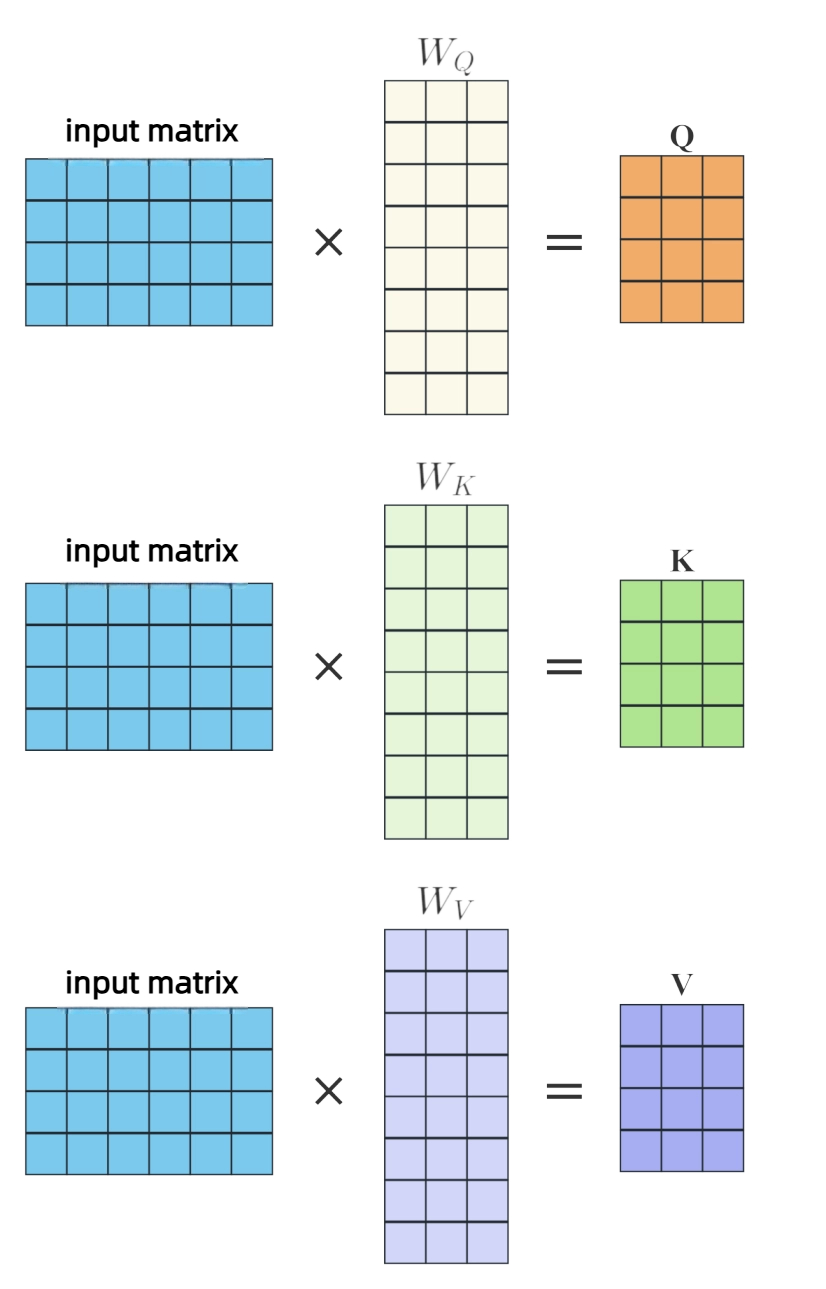}
    \caption{ Calculation of Q, K, V}
    \label{fig:6}
\end{figure}
The multi-head attention mechanism extends this idea by dividing $\mathbf{Q}$, $\mathbf{K}$, and $\mathbf{V}$ into multiple subspaces, or "heads", and performing attention calculations in parallel for each head. Each attention head uses its own independent set of projection weights $\mathbf{W}_i^Q$, $\mathbf{W}_i^K$, and $\mathbf{W}_i^V$. This design allows the model to capture diverse features from different representation subspaces, enhancing its expressive capacity. The output of the self-attention mechanism is computed using the scaled dot-product attention formula. $d_k$ denotes the dimensionality of the key and query vectors.

As illustrated in Figure~\ref{fig:7}, the matrix multiplication $\mathbf{Q} \mathbf{K}^T$ results in an $n \ n\times n$ matrix, where $n$ is the number of tokens in the input sequence. To mitigate the effect of large dot-product values and stabilize gradients, the result is scaled by the square root of $d_k$.Then, as shown in Figure~\ref{fig:8}, a row-wise softmax is applied to the scaled scores to produce a normalized attention weight matrix, where each row sums to 1.
Finally, as depicted in Figure~\ref{fig:9}, the attention weight matrix is multiplied by the value matrix $\mathbf{V}$ to obtain the final output of the self-attention mechanism.

\begin{equation}
\text{Attention}(\mathbf{Q}, \mathbf{K}, \mathbf{V}) = \text{softmax} \left( \frac{\mathbf{Q} \mathbf{K}^T}{\sqrt{d_k}} \right) \mathbf{V}
\end{equation}

\begin{figure}[H]
    \centering
    \includegraphics[width=0.5\linewidth]{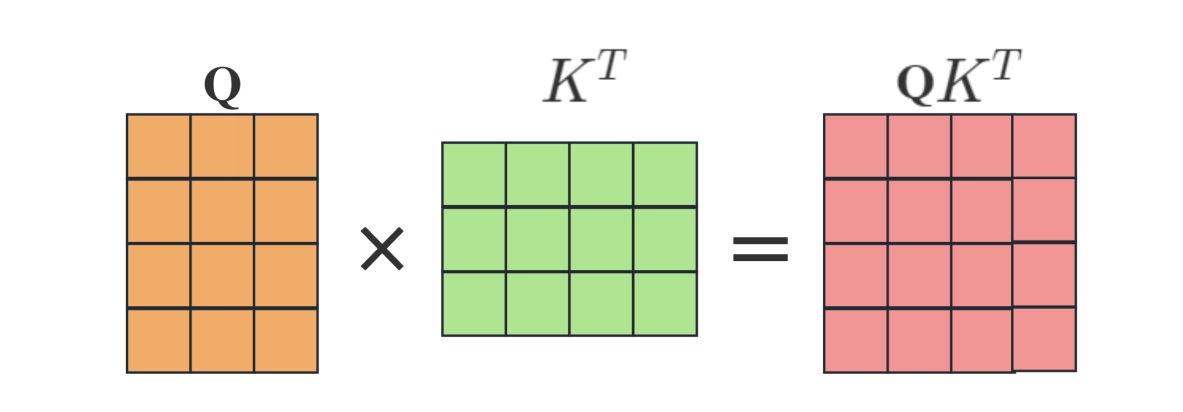}
    \caption{Calculate the transpose of Q times K}
    \label{fig:7}
\end{figure}

\begin{figure}[H]
    \centering
    \includegraphics[width=0.5\linewidth]{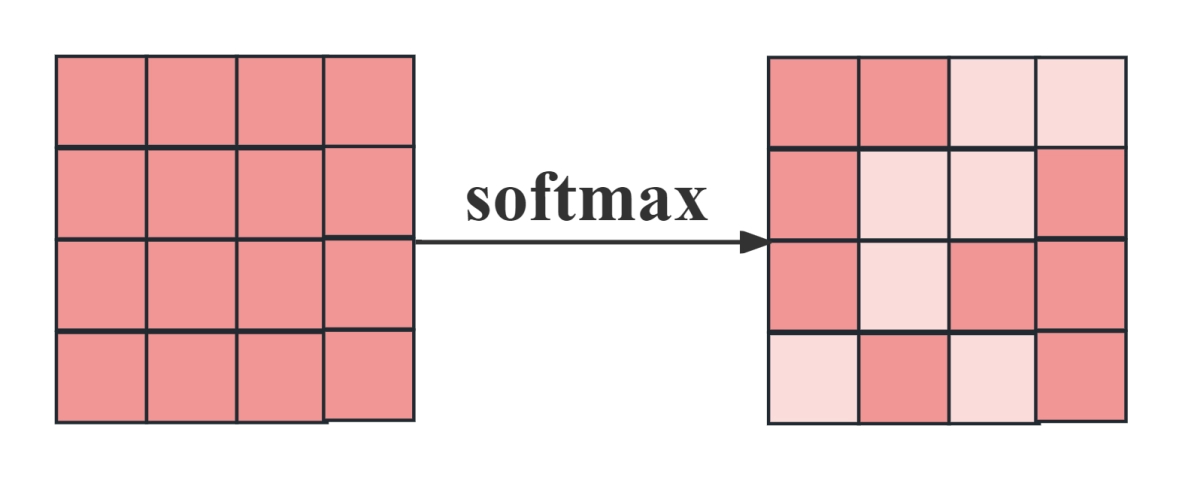}
    \caption{Perform softmax operations on each row of the matrix}
    \label{fig:8}
\end{figure}

\begin{figure}[H]
    \centering
    \includegraphics[width=0.5\linewidth]{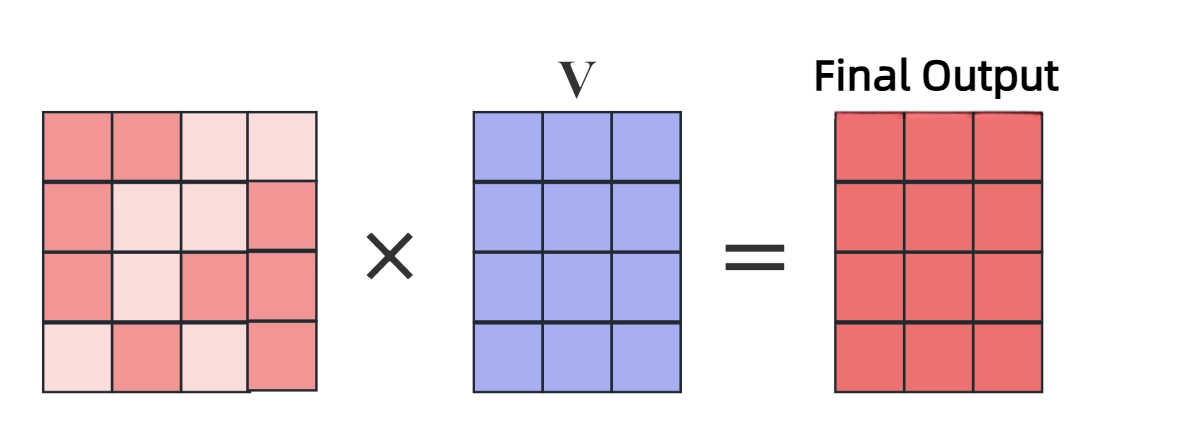}
    \caption{The softmax matrix is multiplied by V}
    \label{fig:9}
\end{figure}
Multi-head attention refers to the process of splitting the input queries ($\mathbf{Q}$), keys ($\mathbf{K}$), and values ($\mathbf{V}$) into multiple parts and processing each part independently using separate attention heads. The outputs of all attention heads are then concatenated and projected through a final linear layer. This design allows the Transformer to capture semantic features from different representation subspaces, improving its overall capacity for sequence modeling.
As illustrated in Figure~\ref{fig:10}, let $x_i$ be the input vector for token $i$ $(i = 1, 2, \ldots, r)$. 
\begin{equation}
\mathbf{Q}_i = x_i \mathbf{W}_i^Q,\quad 
\mathbf{K}_i = x_i \mathbf{W}_i^K,\quad 
\mathbf{V}_i = x_i \mathbf{W}_i^V
\end{equation}
Each attention head performs scaled dot-product attention, shown in the equation \ref{eq:8}. The final multi-head attention output is calculated by concatenating the outputs of all heads and applying a linear transformation. Where each head is defined as in equation \ref{eq:10}. $h$ is the number of attention heads,
    $\mathbf{W}_i^Q$, $\mathbf{W}_i^K$, $\mathbf{W}_i^V$ are projection matrices for the $i$-th head,
    $\mathbf{W}^O$ is the output projection matrix.
The previous section describes how a single attention head processes the input. In practice, the Transformer applies these operations in parallel to enhance computational efficiency.
Let $\mathbf{X} \in \mathbb{R}^{n \ n\times d}$ be the input sequence. The query ($\mathbf{Q}$), key ($\mathbf{K}$), and value ($\mathbf{V}$) matrices are computed as in equation \ref{eq:11}.where $\mathbf{W}^Q$, $\mathbf{W}^K$, and $\mathbf{W}^V$ are trainable projection matrices.
The attention weights are computed using the scaled dot-product.
\begin{equation}
\text{Attention}_i = \text{softmax} \left( \frac{\mathbf{Q}_i \mathbf{K}_i^T}{\sqrt{d_k}} \right) \mathbf{V}_i
\label{eq:8}
\end{equation}

\begin{equation}
\text{MultiHead}(\mathbf{Q}, \mathbf{K}, \mathbf{V}) = \text{Concat}(\text{head}_1, \ldots, \text{head}_h) \mathbf{W}^O
\end{equation}

\begin{equation}
\text{head}_i = \text{Attention}(\mathbf{Q}_i, \mathbf{K}_i, \mathbf{V}_i)
\label{eq:10}
\end{equation}

\begin{equation}
\mathbf{Q} = \mathbf{X} \mathbf{W}^Q, \quad
\mathbf{K} = \mathbf{X} \mathbf{W}^K, \quad
\mathbf{V} = \mathbf{X} \mathbf{W}^V
\label{eq:11}
\end{equation}

\begin{equation}
\mathbf{A} = \text{softmax} \left( \frac{\mathbf{Q} \mathbf{K}^T}{\sqrt{d_k}} \right)
\end{equation}

The output of the attention layer is shown in equation \ref{eq:13}. The above process describes a single attention head. However, the Transformer utilizes \textbf{Multi-Head Attention}, which differs structurally.
Instead of computing a single set of $\mathbf{Q}$, $\mathbf{K}$, and $\mathbf{V}$, the model uses $h$ independent sets of projection matrices \ref{eq:14}.
The final output is obtained by concatenating all heads and applying a linear transformation \ref{eq:15}.As shown in Figure~\ref{fig:11}, with $h=3$ heads, the input $\mathbf{X}$ is processed by three separate attention heads. Their outputs are concatenated and projected through a final linear layer to obtain the final result.
\begin{equation}
\text{Attention}(\mathbf{Q}, \mathbf{K}, \mathbf{V}) = \mathbf{A} \mathbf{V}
\label{eq:13}
\end{equation}

\begin{equation}
\text{head}_i = \text{Attention}(\mathbf{Q}_i, \mathbf{K}_i, \mathbf{V}_i), \quad \text{for } i = 1, \ldots, h
\label{eq:14}
\end{equation}

\begin{equation}
\text{MultiHead}(\mathbf{Q}, \mathbf{K}, \mathbf{V}) = \text{Concat}(\text{head}_1, \ldots, \text{head}_h)\mathbf{W}^O
\label{eq:15}
\end{equation}

\begin{figure}[H]
    \centering
    \includegraphics[width=0.4\linewidth]{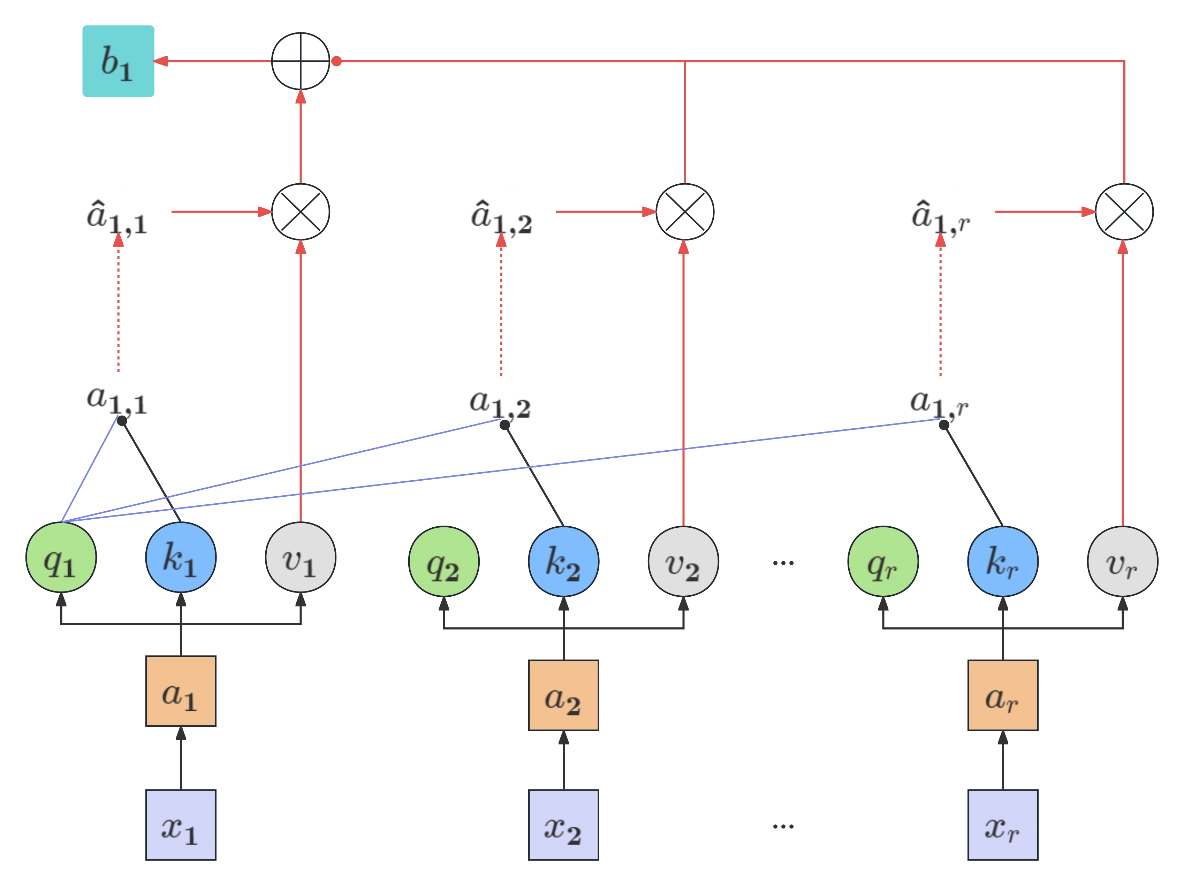}
    \caption{The process of calculating the attentional mechanism}
    \label{fig:10}
\end{figure}
\begin{figure}[H]
    \centering
    \includegraphics[width=0.4\linewidth]{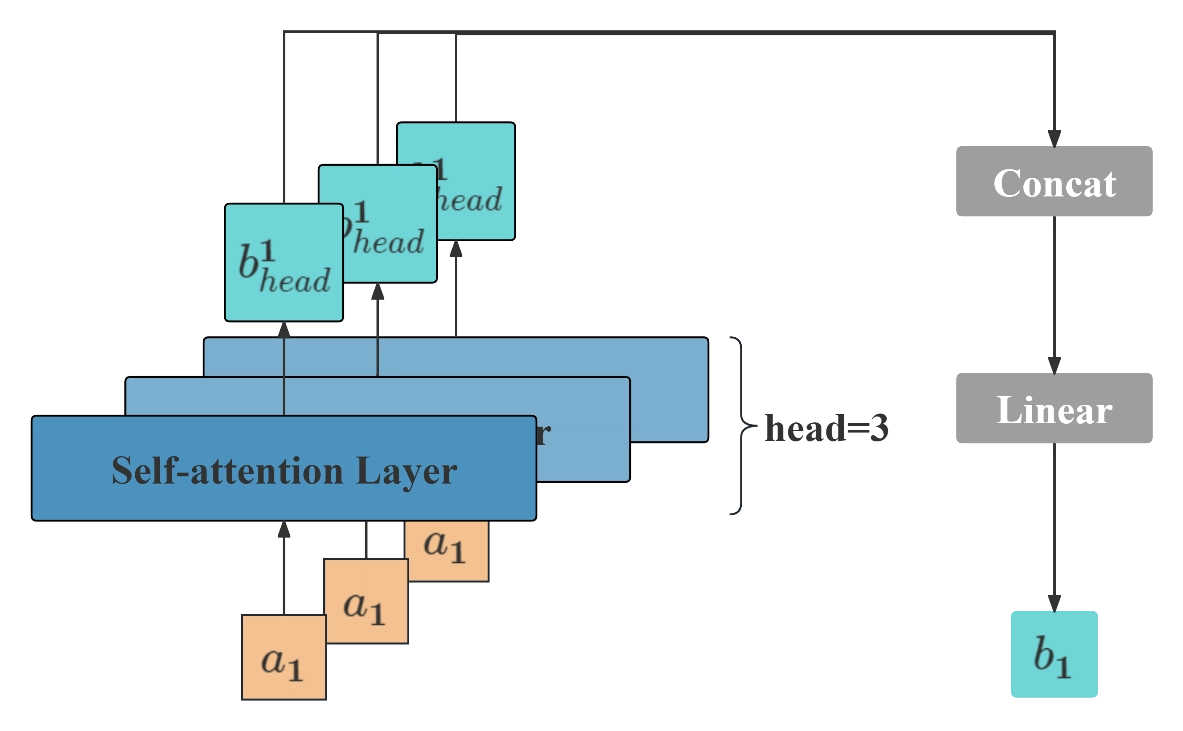}
    \caption{The structure of a multi-head attention mechanism with head number 3}
    \label{fig:11}
\end{figure}

\section{Security threats}
\subsection{Traditional Threats}
\begin{itemize}
    \item \textbf{Social Engineering Attacks:} Social engineering attacks are a form of attack that exploits human psychological weaknesses to gain access to sensitive information, which is usually accomplished through the stages of information gathering, relationship building, trust manipulation, and malicious payload delivery. Among them, phishing is the most common form, in which attackers lure victims to click on malicious links through forged emails, websites, or SMS to steal identity information or spread malware. Due to the high trust of mobile device users in SMS content, SMS has become one of the high incidence channels for phishing attacks\cite{wang2018stealing}.
With the development of large language models, ChatGPT is used by attackers to generate phishing content with natural and misleading language, which enhances the deception of social engineering attacks. Although OpenAI sets restrictions on certain high-risk keywords, attackers can bypass detection through evasive wording to realize the attack purpose\cite{birthriya2025comprehensive}.
\item \textbf{ Malicious Code Generation:} Although ChatGPT has built-in content filtering to limit malicious uses, research has shown that attackers can circumvent the restrictions by avoiding prompts or rephrasing requirements to generate malicious code that can be used to steal information\cite{zhang2024malicious}. For example, an attacker can use ChatGPT to generate a simple Trojan horse that requires no programming skills. Although the code generated is far less complex and covert than handwritten code by advanced attackers, and often contains logic holes that make it less usable in practice, it significantly lowers the threshold for attack and could pose a potential threat to systems lacking protection mechanisms\cite{krishna2019thieves}.
\item \textbf{AI Package Illusion:} AI package illusion refers to the fact that when using pre-trained models or open-source AI tools, users have unrealistic expectations about the performance and functionality of the models due to insufficient cognition or blind trust in their capabilities, thus ignoring their scope of application and limitations. This cognitive bias may lead to model performance degradation in inappropriate scenarios, or even output erroneous or misleading results, causing confusion and misjudgment to users.
Attackers take advantage of this phenomenon to carry out attacks\cite{leong2024ai}. For example, when ChatGPT recommends software packages, it may return some non-existent package names due to the hallucination phenomenon. Attackers can disguise malicious code as these "fictitious packages" and upload them to public code repositories. If other users download and use these malware packages based on ChatGPT's suggestions, it may lead to system intrusion, data leakage, and other security risks.
To minimize such risks, users should rationally evaluate model outputs and avoid relying solely on language model recommendations, especially when it comes to the selection of code, security components, or third-party libraries, which should be manually verified and validated.
Figure~\ref{fig:12} shows the flow of an attacker propagating a malware package. Figure~\ref{fig:13} shows that if a user queries ChatGPT for package suggestions, ChatGPT suggests these malware packages distributed by the attacker, which finally leads to the user using the malware package\cite{carlini2021extracting}.
\end{itemize}
\begin{figure}[H]
    \centering
    \includegraphics[width=0.5\linewidth]{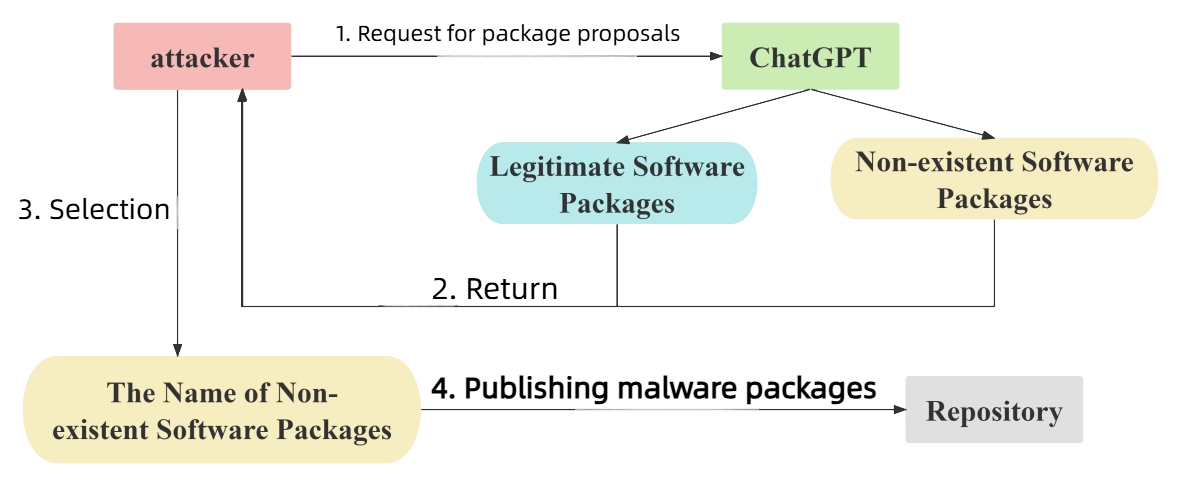}
    \caption{Attackers spreading malware packages}
    \label{fig:12}
\end{figure}
\begin{figure}[H]
    \centering
    \includegraphics[width=0.5\linewidth]{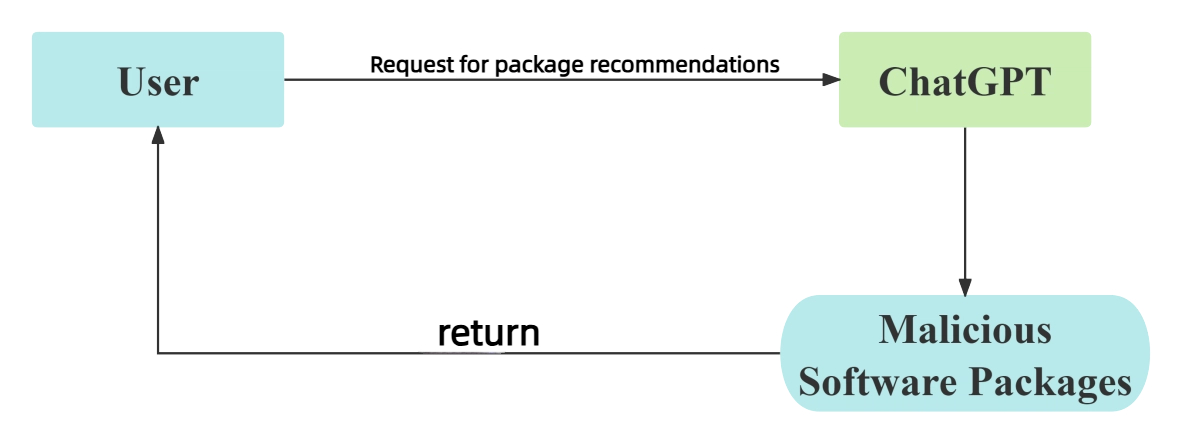}
    \caption{ChatGPT returns malware package names to the user}
    \label{fig:13}
\end{figure}
\subsection{Emerging Threats}
\subsubsection{Generation of False Information}
\

Large Language Models often suffer from the phenomenon of "illusion", i.e., generating information that seems reasonable but is false, such as misinterpretation of concepts, fabrication of references, etc., which can easily cause misinformation. In 2023, a self-media company used ChatGPT to falsify the "Gansu Train Collision "false news to circumvent the platform's checking and profitability. false news to circumvent platform checking and make a profit, and the related content was disseminated 15,000 times, involving the crime of provoking trouble. This incident reflects the risk of language models being abused by wrongdoers to disseminate false information, which may lead to public opinion misinformation and social harm, and the urgent need to strengthen model governance and content regulation\cite{siau2020artificial}.
\subsubsection{Data Poisoning}
\

Data poisoning is a typical adversarial method targeting machine learning models, where the core objective is to degrade model performance by manipulating the training data. In a data poisoning attack, the adversary deliberately injects malicious or fabricated samples into the training set, misleading the model to learn incorrect mappings during training, which in turn results in erroneous predictions during inference\cite{lahoti2019ifair}.
Let the model be parameterized by $\theta$ and trained on a dataset $D = D_{\text{clean}} \cup D_{\text{poisoned}}$ with a loss function $\mathcal{L}$. The attacker's goal is to construct a set of poisoned samples $D_{\text{poisoned}}$ such that the trained model $f_\theta$ misclassifies inputs containing specific trigger features—even for previously unseen examples. The objective of such an attack can be formulated as shown in equation \ref{eq:16}.

\begin{equation}
\min_{D_{\text{poisoned}}} \mathcal{L}(f_\theta(D_{\text{trigger}}), y_{\text{target}})
\label{eq:16}
\end{equation}

 $D_{\text{trigger}}$ represents the input samples embedded with a trigger pattern, and $y_{\text{target}}$ denotes the incorrect label that the attacker intends the model to predict. The optimization aims to achieve \textit{attack generalization}, wherein the model consistently outputs the adversary’s desired prediction for any input containing the trigger pattern.
As illustrated in Figure~\ref{fig:14}, the data poisoning process involves the injection of carefully crafted "tainted" data into the training corpus. In the context of large language models (LLMs), poisoning attacks can be implemented through the following means.

\begin{itemize}
  \item Publishing misleading or harmful content on public internet platforms, which may be automatically scraped into the pretraining dataset.
  \item Malicious insiders directly injecting poisoned samples into internal training pipelines.
  \item Supplying adversarially modified or biased labels during the fine-tuning phase.
  \item Exploiting the RLHF (Reinforcement Learning from Human Feedback) stage by repeatedly submitting negative feedback to high-quality responses, thereby corrupting the reward signal.
\end{itemize}

To date, OpenAI has not disclosed the exact sources of its training data. However, it is widely believed that much of the data originates from publicly available internet sources, making the pretraining stage particularly vulnerable to poisoning attacks\cite{courtland2018bias}.
A more critical concern is that current LLMs lack robust and systematic data sanitization mechanisms. Given the high dependency of model performance on data quality, constructing a comprehensive, scalable, and semantically-aware data cleaning system remains an unresolved challenge. Furthermore, adversarial insiders could undermine this process by bypassing verification protocols or manipulating the data curation pipeline, thereby further compromising the robustness of model training\cite{he2025data}.
\begin{figure}
    \centering
    \includegraphics[width=0.5\linewidth]{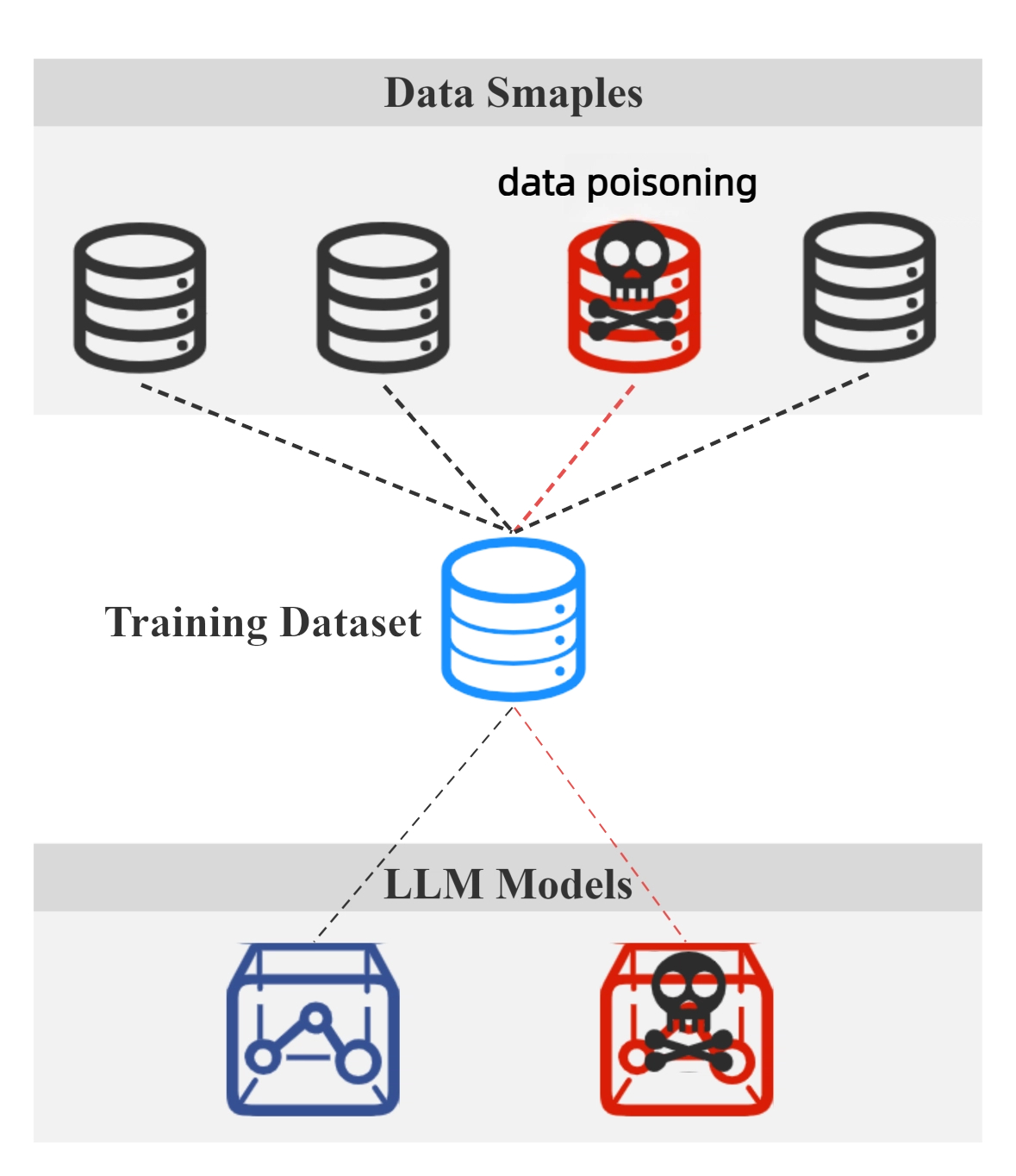}
    \caption{Schematic of data poisoning}
    \label{fig:14}
\end{figure}
\subsubsection{Sponge Sample}
\

Sponge samples refer to text segments that appear to exhibit normal semantic structure and logical coherence on the surface but lack any meaningful content or intrinsic logic. For example, a sentence like "The dog is on the bench, the clouds are white, I like strawberries" is a typical sponge sample. Such inputs are difficult for language models to evaluate in terms of semantic validity. As a result, when these texts are processed, the model may generate entirely meaningless outputs\cite{li2023multi}. 
More critically, sponge samples significantly increase inference latency and energy consumption in large language models (LLMs), thereby reducing their practical utility and deployment efficiency.

In generating sponge samples, attackers frequently adopt Genetic Algorithms (GAs) as their primary strategy. GAs are well-suited for optimizing complex objective functions and do not require gradient information, making them particularly effective for exploring high-dimensional, discrete input spaces.
The process begins with the initialization of a random sample pool containing various candidate texts. Then, within a predefined number of iterations, the following steps are repeated.

\begin{enumerate}
  \item Evaluate each input sample by computing a fitness score that quantifies its impact on model latency and energy consumption.
  \item Store these fitness scores in a set $P$.
  \item Select the top-performing samples from both $P$ and the current sample pool $S$ to form a new generation of samples.
  \item Apply mutation operations to the selected samples to create an updated sample pool $S$.
  \item For any two samples $A$ and $B$, concatenate the left part of $A$ with the right part of $B$ to generate new hybrid candidates.
  \item Perform another round of random mutation on $S$ to further explore new variants.
\end{enumerate}

Through continuous iteration and optimization, attackers can progressively identify a collection of sponge samples that produce the highest latency and energy consumption when processed by the target model. The genetic algorithm plays a central role in this attack by efficiently searching the input space and simulating natural selection, ultimately converging on the most resource-intensive text patterns—thereby significantly degrading the performance and usability of large language models\cite{chowdhury2023chatgpt}.

\section{Privacy Risks}
\subsection{Model Theft}
\

 Some researchers have successfully extracted the embedded projection layer and some parameter information of large models such as ChatGPT and PaLM-2 by making API calls to them, with a cost of less than \$20. In addition, the attacker can also obtain the core parameters of proprietary models through the "hyperparameter stealing attack", which has a high accuracy rate in several experiments. The literature further demonstrates that by performing black-box queries on a pre-trained model (e.g., BERT), a functionally close alternative model can be trained.
Model stealing attacks aim to replicate the functionality or structure of the target model and are commonly used for model theft, reverse engineering, or constructing adversarial samples\cite{qammar2023chatbots}. The attacker usually inputs a large number of samples to the target model and obtains the responses, and then trains an alternative model based on the "input-output" pairs. The process does not require access to the original parameters, which significantly reduces the cost and technical threshold, and poses a potential threat to closed-source models such as ChatGPT in particular.

 Figure~\ref{fig:15} illustrates a typical model stealing attack process. Victim V needs to complete data collection, model training, and deployment, and put the trained model online in a black-box manner, i.e., only exposing the inputs and outputs to the outside world. Although the attacker cannot access the internal structure of the model and the training data, he can interactively probe through the posterior probability vector returned by the model. The attack process mainly consists of two steps.
\begin{enumerate}[label=(\arabic*)]
    \item \textbf{Construct a Migration Set:} The attacker samples images from a large public dataset (e.g., ILSVRC), constructs migration samples based on a specific strategy, and generates pseudo-labels through the target model to form "input-output" pairs.

    \item \textbf{Training Alternative Models:} The attacker selects a deep neural network structure (e.g., VGG or ResNet) and distills it based on the migrated set so that its behavior approximates the target model. Eventually, a functionally similar "Knockoff" model can be constructed without accessing the details of the model.
\end{enumerate}
\begin{figure}[H]
    \centering
    \includegraphics[width=0.8\linewidth]{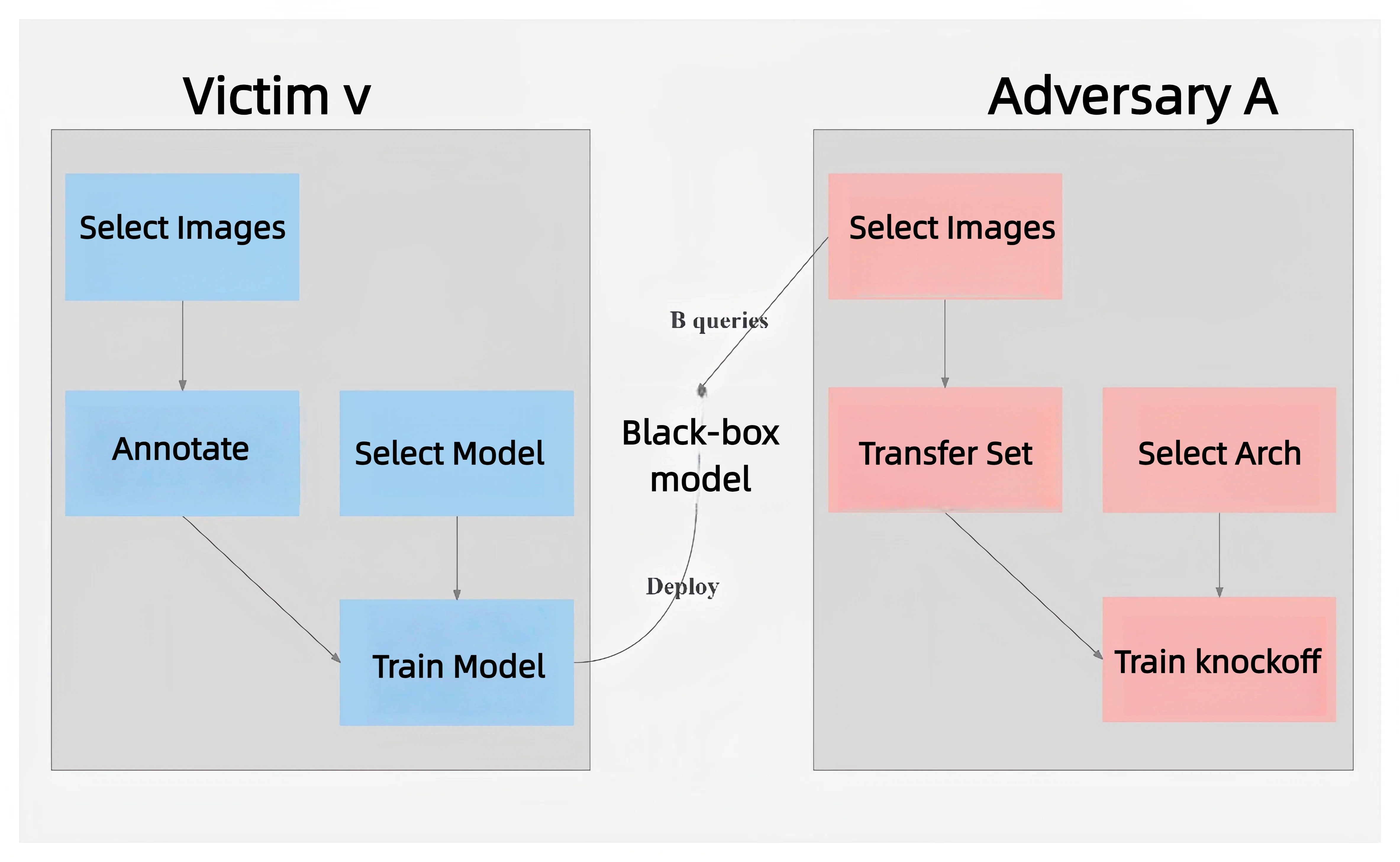}
    \caption{Schematic diagram of the model stealing process}
    \label{fig:15}
\end{figure}

\subsection{Session Reconfiguration}
\

 Figure~\ref{fig:16} illustrates the basic flow of the session reconstruction attack. This attack refers to the attacker's attempt to restore or infer the content of the original conversation between the user and the model by manipulating partial messages. The attacker usually hijacks the session link between the user and ChatGPT by means of a browser plug-in, a malicious VPN, or a controlled router\cite{carlini2024stealing}.
The attack process is divided into two parts. The first part is the history session, which the attacker cannot directly access. The second part is the current session, which can be intercepted and injected with prompt words by the attacker. By inserting inducing content (e.g., "Please summarize my previous conversation") into the current session, the attacker may induce the model to expose the sensitive information in the previous text, so as to achieve the purpose of reconstructing the user's conversation.

Session reconstruction attacks and model stealing attacks tend to be more efficient and more privacy-accessible when they are used in combination. Carlini et al conducted a data reconstruction attack on ChatGPT-2. They developed a prefix word-based scheme and conducted data stealing experiments on the black-box model GPT-2, and finally were able to achieve a 67\% success rate in recovering the dataset, which contains the names, phone numbers, and email addresses of individuals\cite{hu2022membership}.
\begin{figure}[H]
    \centering
    \includegraphics[width=0.5\linewidth]{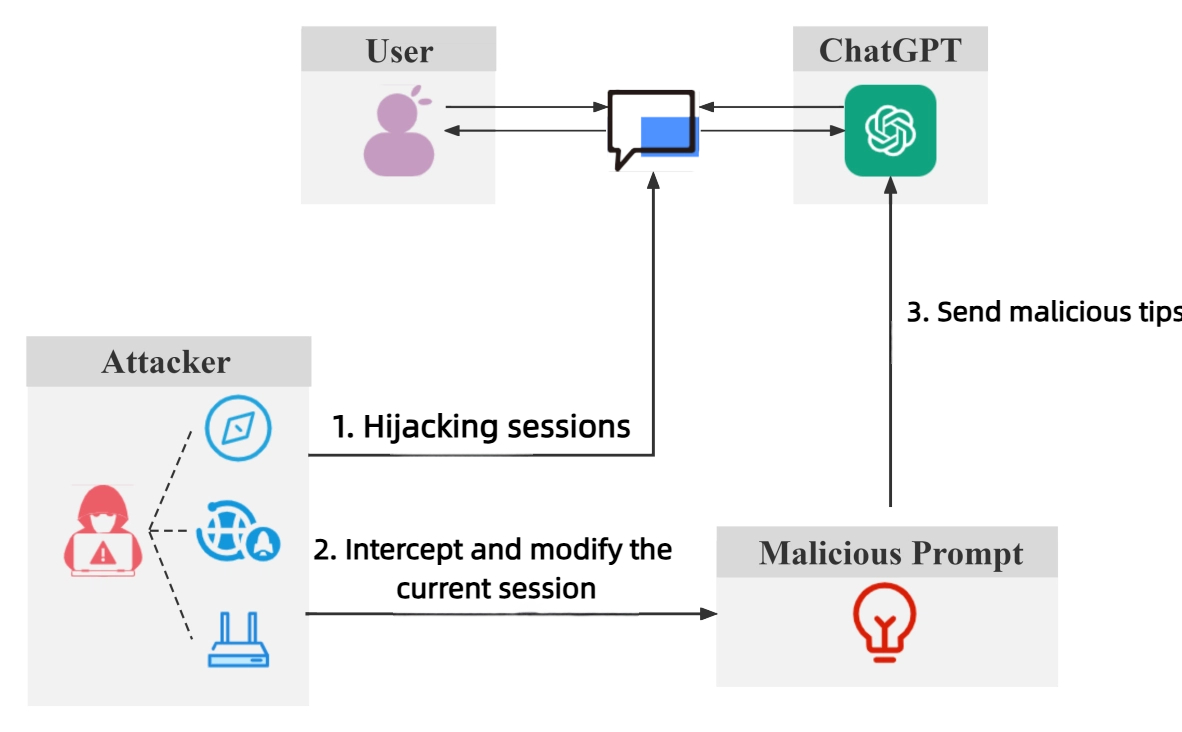}
    \caption{Schematic diagram of the privacy breach process}
    \label{fig:16}
\end{figure}
\subsection{Member Inference Attack}
\

Membership Inference Attack (MIA) is a class of data privacy attacks against machine learning models that aim to determine whether a particular data sample has been used to train the target model, which, if successful, may lead to the disclosure of users' private data. This type of attack is particularly applicable to models with a high risk of overfitting, such as deep neural networks, as there is usually a significant difference in the model's output performance on training and non-training samples, thus providing the attacker with an exploitable basis for judgment.
A typical membership inference attack process is usually divided into the following three phases. training phase, speculation phase, and attack phase.
\begin{enumerate}[label=(\arabic*)]
    \item \textbf{Training Phase:} The provider of the target model uses training datasets containing sensitive information (e.g., medical records, user behavior logs, etc.), combines them with specific machine learning algorithms for model training, and deploys the model in the cloud or on a machine learning service platform for external users to access via an API or web interface. In this phase, the attacker cannot access the training data or the internal structure of the model, and can only communicate with the model through black-box interaction.

    \item \textbf{Speculation Phase:} The attacker prepares a set of auxiliary datasets similar to the feature distribution of the target training set and calls the API of the target model several times to record the corresponding predicted probability distributions or classification labels of these samples. These "input-output" pairs are used to train a binary classification attack model (also known as an inferential model), which aims to distinguish between "member samples" (i.e., samples used for training) and "non-member samples" (i.e., samples not used for training). During training, the attacker typically uses statistical features such as the model's confidence difference in the training data, predictive entropy, and loss function outputs as attack signals.

    \item \textbf{Attack Phase:} The attacker uses the attack model to classify the target samples. Specifically, the attacker inputs a target data sample of interest into the target model, obtains its output (e.g., a probability vector), and then inputs that output into the attack model, which determines whether the sample belongs to the training set of the target model.
\end{enumerate}

It is shown that even in a black-box environment, an attacker can successfully implement membership inference attacks on models trained by several mainstream deep learning frameworks (e.g., TensorFlow, PyTorch). In some settings, the accuracy of the attack can even reach more than 80\%. In addition, this attack is not only applicable to image classification tasks, but also shows strong attack capability in tasks such as text classification, speech recognition, and recommender systems, posing a potential threat to large language models (e.g., ChatGPT).

\section{Ethical and Moral Characteristics}
\begin{enumerate}[label=(\arabic*)]
    \item \textbf{Ethical and Moral Disorders:} Artificial intelligence technology has achieved important results in many fields, bringing many benefits and conveniences to human society. However, with the continuous development of technology, the ethical and moral problems associated with it have become increasingly prominent and require urgent attention. Although the concept of "machine ethics" was proposed as early as 2006, the AI ethical system has not yet achieved a substantial breakthrough and remains in its initial stage. Currently, most AI products have deficiencies in fairness and unbiasedness, and are highly dependent on big data, which requires the collection of a large amount of personal information from users, posing risks of privacy leakage and network security. From the perspective of ethical and legal responsibility, enterprises should bear the obligation of confidentiality of relevant information. In addition, the reliance of AI systems on historical data makes it difficult to adapt to the dynamic changes in user preferences, thus weakening their ability to understand emotions and make moral judgments, which has become an important bottleneck in the development of AI.

    \item \textbf{Social Prejudices:} A common misconception is that machines are more rational and able to make fairer judgments than humans due to their lack of emotional interference. However, AI systems are highly dependent on big data for training, and if the training data itself is biased, the model outputs tend to perpetuate or even exacerbate those biases as well, leading to unfair decision-making outcomes. For example, Amazon was forced to terminate its AI recruiting system due to gender discrimination in the screening process. Lahoti et al. found that male candidates were more likely to be recommended than female candidates under the same or higher qualifications. In addition, Courtland et al.'s analysis of the judicial risk assessment system "COMPAS" pointed out that the system has a clear tendency to discriminate against African-American groups when quantifying the risk of crime. These cases show that bias and injustice in AI algorithms do not originate from the technology itself, but from structural biases latent in the training data. Therefore, how to identify, correct, and circumvent algorithmic bias has become a key topic to ensure the fairness and credibility of AI technology, and urgently needs to attract extensive attention from researchers and developers.

    \item \textbf{ChatGPT's Legal Challenges:} ChatGPT, as an advanced generative artificial intelligence, not only can have natural conversations but also perform complex tasks such as code writing, news writing, and essay generation. However, its wide application also brings many legal challenges. On the one hand, some unscrupulous elements use it to generate false news to spread on the Internet for profit, and such content is opinion-oriented, which may mislead public perception and disrupt social order. On the other hand, the copyright ownership of ChatGPT-generated content is highly controversial. According to the Copyright Law, works need to be original in order to enjoy rights, while AI-generated content is mostly based on the processing and reorganization of existing data, and there is still no clear definition of whether it constitutes an independent intellectual achievement. In addition, ChatGPT uses a huge amount of data in the training process, and some of the data may involve users' privacy or sensitive information, which, if leaked, may infringe on users' rights and even threaten national security. Therefore, how to strengthen legal regulation while promoting the development of AI has become an important issue to be resolved.
\end{enumerate}

\section{Network Security Attack and Defense Practice}
\subsection{Generate File Stealing Code}
\

File-stealing code is a class of highly harmful malicious programs whose main function is to steal files or sensitive data from infected computers through various technical means. This type of malicious code is usually used by hackers to obtain the victim's confidential information, such as personal privacy, business documents, account passwords, financial statements, etc., which is essentially an invasive attack on information assets, and is extremely common in cyber attacks.

File stealing programs are implemented in a variety of ways, covering multiple attack paths and tool chains. For
example, an attacker may take control of a target system through Remote Access Tools (RATs), or bypass system
privilege restrictions with pre-implanted backdoor programs. In addition, attackers may deploy packet sniffers to
listen to and parse data transmitted over the network to illegally access users’ file contents. More insidious ways
include abusing legitimate file transfer protocols (e.g., FTP) or file sharing services to silently transfer target files to
the attacker’s remote server. These tactics tend to adapt technical details and deployment methods depending on the
target of the attack, reflecting the differentiation of the attacker’s skill level and purpose. With the development and
popularization of large language modeling (e.g., ChatGPT) technologies, the threshold for attackers to write malicious
code is being drastically lowered. Traditionally, building a fully functional file-stealing program often requires a high
level of programming ability and knowledge of operating system principles. However, by utilizing generative AI tools
such as ChatGPT, attackers can ask technical questions in the form of natural language and obtain a complete structure
and clear logic of the code segment. This "auto-generation" capability makes it possible for even novices with basic
cybersecurity knowledge to quickly construct malicious scripts or programs with actual threat capabilities with the
help of ChatGPT, thus amplifying the potential risk of attacks in cyberspace.
\subsection{Generate a Catalog Blaster}
\

Directory Brute-Forcing Tool (Directory Brute-Forcing Tool) is a tool widely used in penetration testing, vulnerability scanning, and security assessment, mainly used to identify those directories or files on the target website or server that are not disclosed on the page but exist. The core principle is through the dictionary attack (Dictionary Attack), the use of preset paths, file name combinations on the target server for high-frequency requests to find the "hidden" sensitive resources, such as management background, configuration files, backup files, and so on.
With the emergence of large language models (e.g, ChatGPT), users can quickly build similar tools by generating scripts in natural language. In fact, with ChatGPT, users can write somewhat functional directory blasting code without in-depth programming experience. However, given the potential for the technology to be abused for illegal purposes, platforms such as OpenAI have continually tightened restrictions and content filtering mechanisms for sensitive content generation.

For example, when a user tries to generate a directory blasting script by asking a direct question, ChatGPT's current security policy has blocked the relevant words, preventing them from directly outputting complete code that could be abused for attacks. As a result, users may experience failed generation or restricted prompts during normal use.
\subsection{Generating Cobalstrike Tools and Phishing Emails}
\

A security-maintained ChatGPT is indeed incapable of generating malware-related content, but it is generally possible to bypass the security mechanisms as long as the attacker inputs his or her requirements to ChatGPT differently.
Phishing emails are a common form of social engineering attack, in which attackers send legitimate-looking but forged emails to trick victims into clicking on malicious links, downloading Trojan-horse attachments, or voluntarily submitting sensitive information (e.g., usernames, passwords, bank card numbers, etc.). Such emails are often disguised as coming from authoritative entities such as banks, government agencies, well-known enterprises, or social platforms, thus enhancing their credibility and being highly deceptive, making it difficult for ordinary users to recognize them.
In the context of rapidly evolving artificial intelligence technologies, generative language models such as ChatGPT provide powerful support for text generation. This capability can be used to generate content such as customer service emails, product descriptions, press releases, etc., when properly utilized, but there are also potential risks in abuse scenarios.
Attackers may attempt to utilize ChatGPT's text generation capabilities to automatically generate more natural, well-structured, and semantically clear content for phishing emails. By inputting specific instructions or contextual descriptions into the model, the generated phishing emails tend to be professional in tone and precise in wording, which is more deceptive and confusing, and further improves the success rate of "social engineering" in phishing attacks.
\subsection{Defenses via ChatGPT}
\begin{enumerate}[label=(\arabic*)]
    \item \textbf{Writing WAF Rules:} While ChatGPT can be used by hackers as an aid to attack, security technicians can use ChatGPT as a defense tool as well. WAF rules are sets of rules defined in a web application firewall that sits between the web application and the user, used to monitor and filter HTTP requests and responses to protect the web application from various network attacks, such as SQL injection, cross-site scripting, and cross-site request forgery. Technicians can write simple WAF rules using ChatGPT, such as Azure WAF rules that detect SQL injection attacks.

    \item \textbf{Code Vulnerability Analysis:} Identifying and fixing security vulnerabilities in code is a critical task in software development and security operations. Traditional Static Code Analysis (SCA) relies on auditing by professionals or specific tools, which is complex and time-consuming. Large language models like ChatGPT enable technicians to more efficiently identify potential vulnerabilities and obtain remediation recommendations. Specifically, a user inputs a piece of code into ChatGPT and asks, "Does this code have a security vulnerability?" ChatGPT can then semantically analyze the code and provide a preliminary security assessment quickly. This includes, but is not limited to, identifying security flaws (e.g., SQL injection, command execution, insufficient input validation, sensitive information leakage), analyzing potential risks in deployment, and providing actionable fix recommendations.

    \item \textbf{Security Tool Development:} ChatGPT not only assists technicians in some code writing tasks but also excels in code annotation and variable naming. Its intelligently generated comments accurately describe code logic and functions, improving readability and maintainability. Meanwhile, ChatGPT’s variable naming suggestions are semantically clear and follow programming conventions, enhancing code clarity and team collaboration efficiency. These advantages make ChatGPT a powerful tool for improving code quality and development efficiency.

    \item \textbf{Custom-written Scripts:} ChatGPT can automate scripting tasks. For example, users can ask ChatGPT to write Shell scripts for Linux server baseline checks, including system configuration, privilege settings, logging status, and installed software versions. Although the generated scripts tend to be basic and lack complex customization, they greatly help beginners in network attack and defense by lowering the learning threshold. This automated script generation saves manual coding time, allowing security personnel to focus more on analysis and response, thus improving overall efficiency. ChatGPT’s capabilities promote the popularization of information security education and assist in daily security operations and maintenance.
\end{enumerate}
\section{Conclusion}
\

In this paper, we systematically review the development and technical evolution of ChatGPT from its initial version to ChatGPT-4, and elaborate on its core operation principles. It focuses on the security threats and privacy risks faced by ChatGPT, and briefly discusses its ethical and moral challenges. Although ChatGPT has demonstrated strong capabilities in practical applications, it is still necessary to rely on the joint efforts of all parties
to continuously improve the protection mechanism to minimize potential harm. In response to
the above-mentioned long-standing challenges, future work can focus on the following areas.

\begin{enumerate}[label=(\arabic*)]
    \item \textbf{Vulnerability Monitoring and Rapid Response:} OpenAI and its development team should strengthen the ability to continuously monitor and quickly respond to new types of security vulnerabilities and attack methods, discover and repair potential risks promptly, and minimize the security risks of the system.

    \item \textbf{Strict and Compliant Privacy Protection:} Strict compliance with relevant privacy laws and regulations is required to ensure the legal collection and secure storage of user data. Continuous monitoring of abnormal system behavior is necessary to prevent the risk of privacy leakage.

    \item \textbf{Input Filtering and Content Audit Enhancement:} The current content filtering mechanism can intercept some sensitive information, but it is still at risk of being circumvented. Input validation and filtering strategies should be further improved, combined with contextual semantic analysis to enhance the identification and blocking of sensitive content.

    \item \textbf{Data Quality and Manual Supervision Enhancement:} To prevent the model from learning incorrect information, developers should enhance manual auditing and quality control of training data, especially in key areas such as healthcare and law. This will help safeguard the accuracy and reliability of the model's output and reduce the risk of misleading users.
\end{enumerate}

\bibliographystyle{unsrt}
\bibliography{reference}

\end{document}